\documentclass[12pt]{iopart}
\usepackage{iopams}
\usepackage{epsfig}
\usepackage{cite}
\begin{document}

\title[GIMP Approach to HMX With Cavities Under Shock]
{Generalized Interpolation Material Point Approach to High Melting
Explosive with Cavities Under Shock}
\author{X. F. Pan, Aiguo Xu, Guangcai Zhang and Jianshi Zhu}
\address{Laboratory of Computational Physics, \\
Institute of Applied Physics and Computational Mathematics, P. O.
Box 8009-26, Beijing 100088, P.R.China} \ead{Xu\_Aiguo@iapcm.ac.cn}

\begin{abstract}
Criterion for contacting is critically important for the Generalized
Interpolation Material Point(GIMP) method. We present an improved
criterion by adding a switching function. With the method dynamical
response of high melting explosive(HMX) with cavities under shock is
investigated. The physical model used in the present work is
 an elastic-to-plastic and thermal-dynamical model with Mie-Gr\"uneissen equation of state.
We mainly concern the influence of various parameters, including the
impacting velocity $v$, cavity size $R$, etc, to the dynamical and
thermodynamical behaviors of the material. For the colliding of two
bodies with a cavity in each, a secondary impacting is observed.
Correspondingly, the separation distance $D$ of the two bodies has a
maximum value $D_{\max}$ in between the initial and second impacts.
When the initial impacting velocity $v$ is not large enough, the
cavity  collapses in a nearly symmetric fashion, the maximum
separation distance $D_{\max}$ increases with $v$.
 When the initial shock wave is strong enough to
collapse the cavity asymmetrically along the shock direction, the
variation of $D_{\max}$ with $v$ does not show monotonic behavior.
Our numerical results show clear indication that the existence of
cavities in explosive helps the creation of ``hot spots''.
\end{abstract}
\pacs{62.50.+p; 62.20.Fe; 81.40.Lm } \vspace{2pc}
\submitto{\JPD}\maketitle

\section{Introduction}
Cavitation phenomena are ubiquitous in nature, ranging from solid,
liquid to plasma. Cavity creation and collapse play a very important
role in a large amount of industrial processes, like erosion of
materials, ignition of explosive, comminution of kidney stones,
etc\cite{Baer1998,B2002,XPZZJPCM}. The cavitation phenomena may
occur in a mesoscopic scale, but generally have a profound influence
on the response in the macroscopic scale. As for explosives, the
cavity collapse may result in jetting phenomena and consequently
lead to a much higher temperature increase of the ``hot spots".
Therefore, it has a potential to start local reaction leading to
partial decomposition or run to full detonation\cite{B2002}. On the
one hand, cavitation acts as a means of increasing the sensitivity
of explosive for ignition, but on the other hand, it represents a
potential safety problem for handling such materials. The interest
in both considerations inspired an amount of theoretical,
experimental and numerical work aiming to study the influence of
cavities to the
explosives\cite{B2002,Bowden1952,Kang1992,Mader1985,Mader1989,Mader1979,Cooper2000,
Bourne1989,Bourner1999,Maillet2003,Holian2002,YQL}.

During the collapse procedure explosive material under consideration
is highly distorted and the boundaries should be tracked, which
proposes a very high requirement to the numerical methods. In the
early studies, conventional Eulerian and/or Lagrangian schemes were
used\cite{Kang1992,Mader1985,Mader1989,Mader1979,Cooper2000}. When
conventional Lagrangian methods are used, the mesh distortion
associated with large deformations reduces the
 accuracy and will probably terminate the computations; ultimately,
 a re-meshing treatment is required, which is not a straightforward
task and generally inefficient. It is known that the conventional
Eulerian code is not convenient to track the boundaries. To
continue, several mixed methods have been proposed to combine the
merits of the two methods and overcome their drawbacks. Among them,
the Arbitrary Lagrange-Euler(ALE) method\cite{XZhang2003} is a
typical one. Several researchers have used the ALE method to study
the collapse of cavities in explosive materials\cite{Albert2003}.
The particle-in-cell (PIC) is a second typical mixed
method\cite{H1964}. It also contributed to the research work on
inhomogeneous plastic-bonded-explosives\cite{Bardenhagen1998}.
Additionally, a number of ``meshless methods'' have also been
developed. To overcome the difficulty due to the distortion of mesh,
the meshless method uses a series of discretized lagrangian points
to construct the shape function.  Some meshless methods, such as the
Smooth Particle
Hydrodynamics(SPH)\cite{Gingold1977,Libersky1996,Swegle1995}, Dual
Particle Dynamics(DPD)\cite{Libersky2003}, etc, have been applied in
the research of cavity collapse in explosive
materials\cite{Hong2004, Libersky2003,Tanaka2005}. It should be
mentioned that the microscopic Molecular Dynamics(MD) has also
brought some new understanding of the physics and chemistry
involved\cite{Maillet2003,Holian2002,YQL}. Yet, even with the most
powerful computers in nowadays, the MD simulation is still far from
reaching the practical spatial and temporal scales of real
experiments\cite{Allen2004}.

The physical models used in previous numerical studies are mainly
fluidic ones. Such models ignore many important characteristics of
solid, like plastic strain, hardening, and effects relevant to the
deformation history. In this study we revisit the Material Point
Method(MPM)\cite{BardenhagenGIMP}, recently developed in the field
of computational solid physics, by presenting a new contact
criterion, and then use it to study the dynamical response of high
melting explosive (HMX) with cavities under shock.

The MPM is a descendant of the PIC extended to solid
mechanics\cite{Sulsky92,Sulsky94, Sulsky95,Sulsky96,PXFGF2007}.
Compared with the above-mentioned methods, the MPM provides a robust
and efficient treatment of large deformation issues and a convenient
framework for modeling contact between large numbers of contacting
bodies. In MPM, each body is discretized by a collection of
Lagrangian material points carrying all information required to
advance the solution; the Eulerian background computational mesh is
used to solve the governing equations and the mesh solution is then
used to update the information on material points. Specifically, at
each time step, the MPM calculations consist of two parts: a
Lagrangian part and a convective one. Firstly, the computational
mesh deforms with the body, and is used to determine the strain
increment and the stresses carried by the particles. Then, the new
position of the computational mesh is chosen (particularly, it may
be the previous one); and the velocity field is mapped from the
material points to the mesh nodes. This method has been applied to
handle engineering problems with large
strain\cite{Wiechowski2004,Coetzee2005} and/or dynamical energy
release rate\cite{Tan2002}, fracture in heterogeneous
material\cite{Joris2005}, dynamics failure\cite{ZChen2002,
ZChen2003}, hypervelocity impact\cite{xzhang2006}, thin
membranes\cite{Allen1999}, granular
materials\cite{Bardenhagen2000,Cummins2002,Bardenhagen1998,Bardenhagen2000-2},
etc. The Generalized Interpolation Material Point(GIMP) method uses
a variational form and a Petrov-Galerkin discretization scheme to
overcome the numerical noise of previous MPM\cite{BardenhagenGIMP}.

We study not only the collapse of cavities but also the deformation
and dynamical response of cavities before collapsing under shock.
These are important for understanding better the ignition of
explosive. The elastic-to-plastic and thermal-dynamics model with
the Mie-Gr\"uneissen equation of state consulting the
Rankine-Hugoniot curve is used to simulate the mechanical and
thermal behaviors of the material. The model used here is more
suitable for simulating the dynamical response of the solid material
under shock than previous ones.

This paper is organized as follows. In Sect. 2 we briefly describe
the generalized interpolation material point method where the
contact algorithm is improved. In Sect. 3 the new scheme is
validated by several benchmark tests and then is used to study
cavitated high melting explosive. Sect. 4 concludes the present
paper.

\section{APPROACH}

\subsection{The Generalized Interpolation Material Point method}

For continuum bodies, the conservation equation for mass is
\begin{equation}
\frac{\mathrm{d}\rho}{\mathrm{d}t} + \rho \nabla \cdot \mathbf{v} =
0. \label{eq-pmp-mass}
\end{equation}
where $\rho$ is the mass density, $\mathbf{v}$ is the velocity,

In GIMP and MPM, the continuum bodies are discretized with $N_p$
material particles. Each material particle carries the information
of position, velocity, temperature, mass, density, Cauchy stress,
strain tensor and all other internal state variables necessary for
the constitutive model. Since the mass of each material particle is
equal and fixed, Eq.(\ref{eq-pmp-mass}) is automatically satisfied.
At each time step, the mass and velocities of the material particles
are mapped onto the background computational mesh(grid). The mapped
nodal velocity $\mathbf{v}_j$ is obtained through the following
equation,
\begin{equation}
\sum_{j}{m_{ij}\mathbf{v}_j} = \sum_{p}{m_p \mathbf{v}_p
N_{i}(\mathbf{x}_p)} \label{eq-pmp-velo}
\end{equation}
where $m_p$, $\mathbf{v}_p$ and $\mathbf{x}_p$ are the mass,
velocity and position of particle $p$, respectively. $N_i$ is the
shape function, $i$ and $j$ indexes of node.

\begin{figure}[tbp]
\centering
\includegraphics*[scale=0.8]{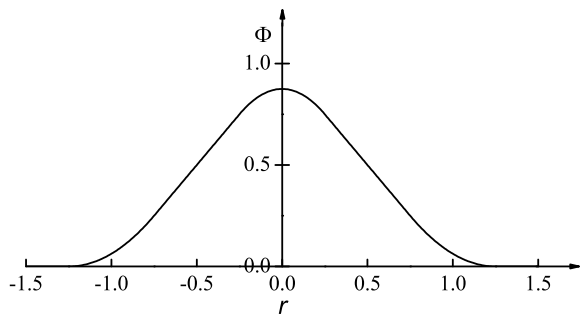}
\caption{Contiguous particle GIMP weighting function in one
dimension.} \label{Fig1}
\end{figure}

In the early version of MPM, the grid shape function $N_i$ is not
smoothed in the construction of the weighting function which causes
the numerical noise as the material points cross computational grid
boundaries. Bardenhangen et al\cite{BardenhagenGIMP} presented a
family of methods named the Generalized Interpolation Material
Point(GIMP) methods in which the interpolation function are in
$C^1$(as opposed to MPM, for which are in $C^0$). In this paper the
following smoothed shape function in $C^1$ is used,
 $N_i=\Phi(r_x)\Phi(r_y)$, where $r_x=|x_p-x_i|/L$,
$r_y=|y_p-y_i|/L$, $L$ is the length of cell, $\Phi(r)$ is given in
flowing,

\begin{equation}
\Phi(r) = \left\{
\begin{array}{ll}
\frac{7-16r^2}{16}, & r \le 0.25 \\
1-r, & 0.25 < r \le 0.75 \\
\frac{(5-4r)^2}{16}, & 0.75 < r \le 1.25 \\
 0,  & r > 1.25
\end{array}
\right. \label{eq-mpm-shape}
\end{equation}
See also Fig.\ref{Fig1}. Eq.(\ref{eq-mpm-shape}) is a weighting
function with support in adjacent cells and the next nearest
neighbor cells. This specialization has the advantage that it
develops weighting function in $C^1$ with a minimal amount of
additional complexity. The increased support does result in an
increase in computational effort that is different from the meshless
methods which invest a mount of effort to the research of the
influence nodes to construct the shape function.

In Eq.(\ref{eq-pmp-velo}), the consistent mass matrix, $m_{ij}$, is
\begin{equation}
m_{ij} = \sum_{p} m_p N_{i}(\mathbf{x}_p) N_{j}(\mathbf{x}_p).
\label{eq-pmp-mass-matrix}
\end{equation}
In practice, we generally replace $m_{ij}$ with a lumped, diagonal
mass matrix so that Eq. (\ref{eq-pmp-velo}) becomes
\begin{equation}
m_{i}\mathbf{v}_i = \sum_{p} m_p \mathbf{v}_p N_i( \mathbf{x}_p )
 \label{eq-pmp-mass-velo-2}
\end{equation}
where lumped mass is
\begin{equation}
m_{i} = \sum_p m_p N_i(\mathbf{x}_p)
 \label{eq-pmp-mass-2}
\end{equation}

The conversation equation for momentum reads,
\begin{equation} \rho
\frac{\mathrm{d}\mathbf{v}}{\mathrm{d}t}=\nabla \cdot \mathbf{\sigma }+\rho \mathbf{b%
}\texttt{,} \label{eq-pmp-strongform}
\end{equation}
where $\mathbf{\sigma}$ is the stress tensor and $\mathbf{b}$ is the
body force. The weak form of Eq.(\ref{eq-pmp-strongform}) based on
the standard procedure used in the FEM\cite{Sulsky94,Sulsky95} can
be written

\begin{equation}
\int_{\Omega}{ \rho \delta \mathbf{v} \cdot
\frac{\mathrm{d}\mathbf{v}}{\mathrm{d}t}\mathrm{d}\Omega} +
\int_{\Gamma_t}{ \delta\mathbf{v} \cdot (\mathbf{\sigma} \cdot
\mathbf{n}-\mathbf{t}) \mathrm{d}\Gamma} + \int_{\Omega}{ \rho
\delta \mathbf{v} \cdot \mathbf{b} \mathrm{d}\Omega } = 0
\label{eq-pmp-weakform}
\end{equation}
where $\mathbf{n}$ and $\mathbf{t}$ is the outward normal unit and
traction vectors on the boundary.

Since the continuum bodies is described with the use of a finite set
of material particles, the mass density can be written as,
\begin{equation}
\rho(\mathbf{x}) = \sum_{p=1}^{N_p}{
M_p\delta(\mathbf{x}-\mathbf{x}_p) } \label{eq-pmp-density}
\end{equation}
where $\delta$ is the Dirac delta function with dimension of the
inverse of volume. The substitution of Eq. (\ref{eq-pmp-density})
into Eq. (\ref{eq-pmp-weakform}) converts the integral to the sums
of quantities evaluated at the material particles, namely,

\begin{equation}
m_{i} \frac{\mathrm{d}\mathbf{v}_i}{\mathrm{d}t} =
(\mathbf{f}_i)^{\mathrm{int}} + (\mathbf{f}_i)^{\mathrm{ext}}
\label{eq-pmp-solve}
\end{equation}
where $m_i$ is the lumped mass and the internal force vector is
given by
\begin{equation}
(\mathbf{f}_i)^{\mathrm{int}} = - \sum_{p}^{N_p} { M_p
\mathbf{\sigma} \cdot (\nabla N_i) / \rho_p } \label{eq-pmp-fint}
\end{equation}
and the external force vector reads
\begin{equation}
(\mathbf{f}_i)^{\mathrm{ext}} = \sum_{p=1}^{N_p}{ N_{i}\mathbf{b}_p
+ \mathbf{f}_i^c  } \label{eq-pmp-fext}
\end{equation}
where the vector $\mathbf{f}_i^c$ is the contact force which is the
external nodal force not including the body force and is illustrated
in the following section.

A explicit time integrator is used to solve Eq. (\ref{eq-pmp-solve})
for the nodal accelerations, with the time step satisfying the
stability condition. The critical time step is the ratio of the
smallest cell size to the wave speed. After the equations of motion
are solved on the cell nodes, the new nodal values of acceleration
are used to update the velocity of the material particles. The
strain increment for each material particle is determined with the
use of gradient of the nodal basis function evaluated at the
material particle position. The corresponding stress increment can
be found from the constitutive model. The internal state variables
can also be completely updated. The computational mesh may be
discarded, and a new mesh is defined, if desired for the next time
step. As a result, an effective computational mesh could be chosen
for convenience\cite{PXFGF2007}.

\subsection{Contact algorithm\label{SEC-CONTACT}}
The GIMP method provides a natural no-slip contact algorithm based
on a common background mesh. But natural contact algorithm has two
drawbacks: firstly, the premature contact occurs because the
velocities of two bodies are mapped on the same nodes though the
distance between the bodies may be still two or even more times of
the length of cell, which may cause the numerical noise of stress;
Secondly, it is impossible to separate the contacting bodies.
Bardenhagen et al have proposed a contact algorithm to simulate the
interactions of the grains of granular
material\cite{Bardenhagen2000}, in which the contact between bodies
is handled when the velocity field of individual particles in
contact differs from the single, center-of-mass velocity field in
the cell containing contacting particles. A multi-mesh mapping
scheme is proposed by Hu and Chen\cite{Hu_Chen2003}. In the
multi-mesh mapping scheme, each material lies in an individual
background mesh rather than in the common background one. The
meshing process of spur gears is simulated by Hu and Chen with their
contact algorithm. To avoid interpenetration and allow separation in
the gear meshing process, the normal velocity of any particle at the
contact surface is calculated in the common background mesh, while
the tangential velocity is found based on the corresponding
information in respective individual mesh.
\begin{figure}[tbp]
\centering
\caption{(See attached 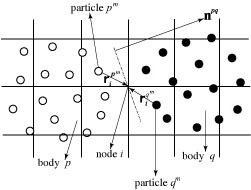)The scheme of contact between body
$p$ and body $q$.} \label{Fig2}
\end{figure}

In Bardenhagen's and Chen's contact algorithms the premature contact
may still occur. In this paper, the criterion of contact consults
the distance of the bodies. In the improved contact algorithm, if
the velocities of body $p$ and body $q$ are mapped on the same node
$i$ (Seen in Fig.\ref{Fig2}), the distance between body $p$ and $q$
is calculated. We note the distance between body $p$ and $q$ as
$\mathfrak{R}_i^{pq}$, which can be calculated as:
\begin{equation}
\mathfrak{R}_i^{pq}= \mathbf{n}^{pq} \cdot
\mathbf{r}_i^p-\mathbf{n}^{pq} \cdot \mathbf{r}_i^q
\label{eq-contact-distance}
\end{equation}
where $\mathbf{n}^{pq}$ is the normal direction of the contacting
interface pointing from body $p$ to $q$. $\mathbf{r}_i^p$ and
$\mathbf{r}_i^q$ are the vectors pointing to node $i$ from body $p$
and body $q$, respectively. $\mathbf{n}^{pq} \cdot \mathbf{r}_i^p$
and $\mathbf{n}^{pq} \cdot \mathbf{r}_i^q$ can be chosen as
\begin{equation}
\mathbf{n}^{pq} \cdot \mathbf{r}_i^p =
\max{\{\mathbf{r}_i^{p^m}\cdot \mathbf{n}^{pq}, m=1,2,...,N_p
\hspace{1mm} \mathrm{and} \hspace{1mm} \mathrm{particle}
\hspace{1mm} p^m \hspace{1mm} \in \hspace{1mm} \mathrm{body}
\hspace{1mm} p\}}
\end{equation}
\begin{equation}
\mathbf{n}^{pq} \cdot \mathbf{r}_i^q =
\min{\{\mathbf{r}_i^{q^m}\cdot \mathbf{n}^{pq}, m=1,2,...,N_q
\hspace{1mm} \mathrm{and} \hspace{1mm} \mathrm{particle}
\hspace{1mm} q^m \hspace{1mm} \in \hspace{1mm} \mathrm{body}
\hspace{1mm} q\}}
\end{equation}
where $\mathbf{r}_i^{p^m}$ and $\mathbf{r}_i^{q^m}$ are the vectors
pointing to node $i$ from particle $p^m$ and $q^m$, $N_p$ and $N_q$
are the numbers of particles belonging to body $p$ and $q$,
respectively.

The the criterion of contact can be written as
\begin{equation}
\mathfrak{R}_i^{pq} \le \frac{L}{2}, \label{eq-contact-criterion}
\end{equation}
where $L$ is the length of cell. If Eq.(\ref{eq-contact-criterion})
is satisfied, the velocities of $p$ and $q$ are adjusted to new
values so that the normal components of them are equal, then the
strain and stress of all the particles are updated. So,
Eq.(\ref{eq-contact-criterion}) plays the role of a switch function.
Once bodies $p$ and $q$ contact, they move together along the normal
until they separate, so the acceleration along the normal of body
$p$ is equal to that of $q$ during the course of the contact. That
is
\begin{equation}
\mathbf{a}_i^p \cdot \mathbf{n}_i^{pq} = \mathbf{a}_i^q \cdot
\mathbf{n}_i^{pq} \label{eq-contact-rate-5}
\end{equation}
where $\mathbf{a}_i^p$ and $\mathbf{a}_i^q$ are the accelerations of
bodies $p$ and $q$ at node $i$, respectively. They can be obtained
from the Newtonian second law. Also the normal contact force
$f_i^{\mathrm{nor}}$ can be derived. Note that the normal contact
force must be nonnegative. So, once $f_i^{\mathrm{nor}}$ is
negative, body $p$ and $q$ is not in contact in the next time step.

If without friction, the contact algorithm has been finished up to
now. In the case with friction the Coulomb friction is applied to
calculate the tangential contact force.

\subsection{Constitutive model and the Equation of State}
In this paper, the material is modeled using von Mises plasticity
with linear harding\cite{XPZZJPCM}. A plastic model dictates a
linear elastic response until a yield criterion is reached. The von
Mises yield criterion is $3J_2-Y^2=0$, where $Y$ is the plastic
yield stress, $J_2$ is the second invariant of $\mathbf{s}$,
$J_2=\frac{1}{2}\mathbf{s}:\mathbf{s}$, $\mathbf{s}$ is the
deviatoric stress tensor,
$\mathbf{s}=\mathbf{\sigma}-\mathrm{Tr}[\mathbf{\sigma}]/3$. The
linear hardening means that $Y$ increases linearly with the second
invariant of the plastic strain tensor. If $3J_2 > Y^2$, the
increment of equivalent plastic strain $\mathrm{d}\varepsilon_p$ can
be calculated as
$\mathrm{d}\varepsilon_p=(\sqrt{3J_2}-Y)/(3G+E_{\mathrm{tan}})$,
where $G$ and $E_{\mathrm{tan}}$ are the shear and harding modulus,
respectively. The increment of the plastic energy can be calculated
as $\mathrm{d}W_p=\mathrm{d}\varepsilon_p \cdot Y$. It is totally
translated into the internal energy.

The pressure $P$ is calculated by using the Mie-Gr\"uneissen state
of equation which can be written as
\begin{equation}
P-P_H = \frac{\gamma(V)}{V}[E-E_H(V_H)] \label{eq-eos}
\end{equation}
This description consults the Rankine-Hugoniot curve. In
Eq.(\ref{eq-eos}), $P_H$, $V_H$ and $E_H$ are pressure, specific
volume and energy on the Rankine-Hugoniot curve, respectively. The
relation between $P_H$ and $V_H$ can be estimated by experiment and
can be written as
\begin{equation}
P_H=\left\{
\begin{array}{ll}
\frac{\rho_0
c_0^2(1-\frac{V_H}{V_0})}{(\lambda-1)^2(\frac{\lambda}{\lambda-1}\times
\frac{V_H}{V_0}-1)^2}, & V_H \le V_0 \\
\rho_0^2 c_0^2 (\frac{V_0}{V_H}-1), & V_H > V_0
\end{array}
\right.
\end{equation}
 where $c_0$ is the sound speed, $\lambda$ is the parameter in the
linear ration, $U_s=c_0 + \lambda U_p $, $U_s$ and $U_p$ are shock
velocity and velocity of particle, respectively.
 In this paper, the coefficient of Gr\"uneissen $\gamma(V_H)$ is
taken as a constant and the transformation of specific internal
energy $E-E_H(V_H)$ is taken as the plastic energy. Both the shock
compression and the plastic work cause the increasing of
temperature. The increasing of temperature from shock compression
can be calculated as:
\begin{equation}
\frac{\mathrm{d}T_H}{\mathrm{d}V_H}=\frac{c_0^2\cdot \lambda
(V_0-V_H)^2}{c_{v}\big[(\lambda-1)V_0-\lambda V_H
\big]^3}-\frac{\gamma(V)}{V_H}T_H. \label{eq-eos-temprshock}
\end{equation}
where $c_v$ is the specific heat. Eq.(\ref{eq-eos-temprshock}) can
be resulted with thermal equation and the Mie-Gr\"uneissen state of
equation\cite{explosion}. The increasing of temperature from plastic
work can be calculated as:
\begin{equation}
\mathrm{d}T_p= \frac{\mathrm{d}W_p}{c_v}\label{eq-eos-temprplastic}
\end{equation}
Both the Eq.(\ref{eq-eos-temprshock}) and the
Eq.(\ref{eq-eos-temprplastic}) can be written as the form of
increment. In the present work the thermal dissipation is not taken
into account. Such a treatment is reasonable for cases where the
propagation speed of shock is much faster than that of the thermal
dissipation.

The material constants are chosen to model the energetic crystal
high melting explosive. The elastic modulo is $11.87$GPa, the
poisson ratio is $0.25$, the density is $1.9\times10^{-3}
\mathrm{g}/\mathrm{mm^3}$, the initial yield strength is 100MPa, the
hardening modulus is 0MPa, the heat capacity is
$1250\mathrm{J/(Kg\cdot K)}$. The coefficient of Gr\"uneissen is
taken as 1.1, $\lambda$ is 2.6 and the sound speed $c_0$ is $2740
\mathrm{m/s}$\cite{Baer1998}.

\section{Results}
In this paper we focus on the two-dimensional case. The shock wave
to material target is loaded via the impacting by a second material
block with symmetric configuration and opposite velocity. The
initial shock is along the vertical direction. We apply periodic
boundary conditions to simulated system in the horizontal direction.
As the first step, we set a single cavity in the HMX block. Such a
simulation model corresponds also to a very wide system with a row
of cavities being parallel to the impacting plane. We study various
cases with different cavity radius and with different strengths of
shock. The influence of existing cavity to the contact of bodies is
investigated and a secondary impacting is observed. In the cases of
strong shock, the influence of cavity size and the impact speed to
the ``hot-spot" is systematically investigated.

\subsection{Validation of the improved contact algorithm\label{sec-validation}}
\begin{figure}[btp] \centering
\includegraphics*[width=0.47\textwidth]{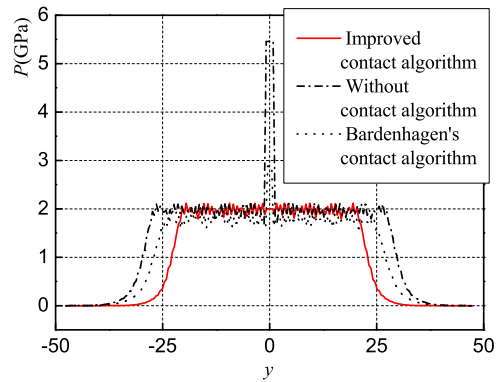}
\caption{(Color online) Pressure distribution along the vertical
direction at time $t=0.014\mathrm{ms}$. The impacting interface is
at $y=0$. Previous contact algorithms present unreasonable too high
pressure in the vicinity of the impact interface.}
\label{Fig_1DPressure}
\end{figure}

The newly proposed contact algorithm is validated by simulating the
impacting of two identical solid bodies having opposite velocities
in vertical direction. The initial speed of each body is set to be
$300\mathrm{m/s}$. Initially, the distance between the two bodies is
set as $4\mathrm{mm}$. The width and height of blocks are set to be
40mm and 50mm, respectively. Fig.\ref{Fig_1DPressure} shows the
pressure along the vertical direction calculated by different
contact algorithms, where the impacting interface is at $y=0$. From
the numerical results we can see that both the GIMP with
Bardenhagen's contact algorithm and the MPM without contact
algorithm give an unreasonable and too high pressure in the vicinity
of the interface corresponding to a severe numerical premature
contact (See the dash-dotted and dotted lines in the figure). The
result by the improved contact algorithm is much better (See the red
line in the figure). It is well-known that spurious ``wall-heating''
phenomenon is generally difficult to eliminate for computational
fluid dynamics and computational solid dynamics.
  Here, our new contact scheme decreases the ``wall-heating'' phenomenon to
an extent that is nearly undistinguishable.

\subsection{Dynamical response to weak shock}
\begin{figure}[btp] \centering
\caption{(See attached 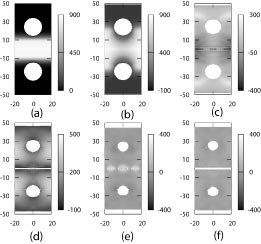)Snapshots for the impact of two
blocks with a single cavity in each. Impact velocity of each block
is $v=150\mathrm{m/s}$. From black to white the gray level in the
figure shows the increase of local pressure. The unit of pressure is
MPa. (a) $t=0.004\mathrm{ms}$, (b) $t=0.008\mathrm{ms}$, (c)
$t=0.014\mathrm{ms}$, (d) $t=0.025\mathrm{ms}$, (e)
$t=0.06\mathrm{ms}$, (f) $t=0.18\mathrm{ms}$.} \label{Fig3}
\end{figure}

Due to the ability of withstanding the deformation of shear and
tension, cavities in the solid material will not collapse when the
shock is very weak. Fig. \ref{Fig3} shows a series of snapshots for
a case with an initial impact speed of 150$\mathrm{m/s}$. The size
of blocks is the same as section \ref{sec-validation} and the radius
of each cavity is 10 mm. The global procedure can be described as
follows:

(a), The two bodies get contact and a shock wave is loaded to each.
The shock waves in the two bodies move forwards to the two cavities.
The pressure of shock is about 900$\mathrm{MPa}$(See
Fig.\ref{Fig3}(a)).

(b), The shocks arrive at the cavities, then rarefactive waves are
reflected back to material in between the two cavities.  The
cavities begin to shrink in a nearly isotropic fashion(See in
Fig\ref{Fig3}(b)).

(c), The rarefactive waves reach the surface of contacting, then the
two bodies begin to separate(See in Fig.\ref{Fig3}(c)). Note that if
there is no cavity inside, the two bodies will not separate so
early.

(d), The two bodies continue to separate until the distance between
the two bodies reaches the maximal value(See in Fig.\ref{Fig3}(d)),
then the two bodies begin to move closer again because the
particles, being above the upper cavity and down the lower cavity,
are still moving towards the contacting surface, which draws the
separating bodies back.

(e), The approaching two bodies collide again. They stay together
until the rarefactive waves reflected from the lowermost and the
uppermost free surfaces arrive at the contacting surface. Then, they
will separate again (See in Fig.\ref{Fig3}(e)).

(f), After e, the two bodies separate. This is the final separation.
The terminal configuration of cavities are still approximately
circular. From the whole procedure, we can see that, due to the
existence of cavities, the two bodies get a secondary impacting.

\begin{figure}[btp]
\centering
\includegraphics*[width=0.47\textwidth]{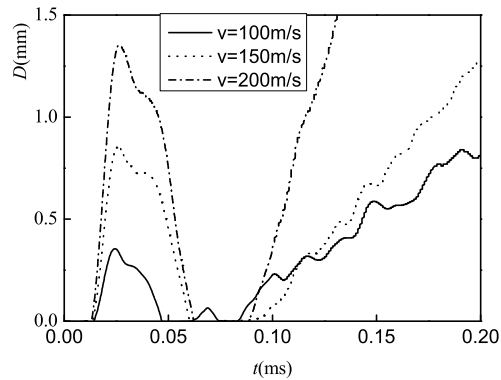}
\caption{The separation distance versus time $t$ for different
impacting speeds shown in the legend. $D$ represents half of the
distance between the two impacting surfaces. Before the occurrence
of the secondary impact, $D$ has a maximum value $D_{\max}$ which
increases with the impacting speed.} \label{Fig_R10_ChangeV}
\end{figure}

In order to comprehend more clearly the phenomena of the secondary
impacting, we alter the impacting speed(in the limit of no collapse)
and the size of cavity. Fig.\ref{Fig_R10_ChangeV} shows the effects
of the initial impacting speed to the maximum separation distance
between the two bodies after the first impacting. In
Fig.\ref{Fig_R10_ChangeV}, $D$ represents half of the distance
between the two impacting surfaces. It varies with time $t$. The
initial impacting speed is set as $100\mathrm{m/s}$,
$150\mathrm{m/s}$ and $200\mathrm{m/s}$, respectively. In all cases
the phenomena of the secondary impacting are observed. The maximum
value of $D$ is denoted by $D_{\max}$. The larger the initial
impacting speed, the larger the value of $D_{\max}$.

\begin{figure}[btp]
\centering
\includegraphics*[width=0.47\textwidth]{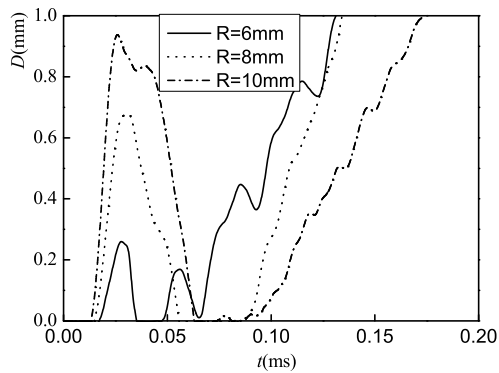}
\caption{The distance of separation versus time $t$ for different
cavity sizes shown in the legend. Here $D$ and $D_{\max}$ have the
same meanings as in Fig.\ref{Fig_R10_ChangeV}.}
\label{Fig_V150_ChangeR}
\end{figure}

Fig.\ref{Fig_V150_ChangeR} shows the effects of cavity size on the
separation distance between the two impacting bodies. Here $D$ and
$D_{\max}$ have the same meanings as in Fig.\ref{Fig_R10_ChangeV}.
The initial radius of each cavity is set as $6\mathrm{mm}$,
$8\mathrm{mm}$ and $10\mathrm{mm}$, respectively. It can be seen
that the larger the radius of cavity, the larger the value of
$D_{\max}$. That is to say, the larger the cavities, the easier to
observe the phenomenon of the secondary impacting. Naturally,  if
the radius of cavity diminishes to zero, the secondary impact
disappears.

\begin{figure}[btp]
\centering
\caption{(See attached 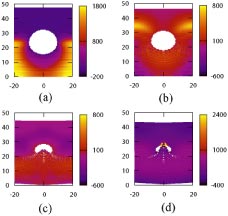) Snapshots for the impacting process
of two blocks with a single cavity. The initial radius of cavity
$10\mathrm{mm}$, impact speed $v=300\mathrm{m/s}$. From blue to red,
the local pressure increases. The unit of pressure is Mpa.
$t=0.008\mathrm{ms}$ in (a), $t=0.012\mathrm{ms}$ in (b),
$t=0.02\mathrm{ms}$ in (c), $t=0.024\mathrm{ms}$ in (d).}
\label{Fig7}
\end{figure}

\subsection{Dynamical response to strong shock\label{sec-strongshock}}
If the shock is strong, asymmetric collapse of the cavity will
occur. When the radius of cavity is set to be 6mm, 8mm and
10mm,respectively, the corresponding minimum impact speed $v$ for
the cavity to collapse asymmetrically is 120m/s, 150m/s and 200m/s,
respectively. That is to say, the smaller the radius of cavity, the
easier for the cavity to collapse asymmetrically. This is because
larger curvature is easier to accumulate particles. The course of
the collapse of the upper cavity can be described as follows: When
the shock reaches the cavity, the upstream side of cavity will
reflect a rarefactive wave which propagates downwards. See
Fig.\ref{Fig7}(a). This causes the pressure in the region close to
the upstream side of the cavity is lower than around. Then, the
neighboring particles accumulate towards to the lower side of
cavity, which will accelerate the upward speed of particles there.
There are two factors that influence the deformation of the cavity.
One is the the accumulation of the particles to the lower side of
the cavity, the other is the resistance of the solid material to
shear. The competition of the two factors determines how the cavity
deforms. Obviously, as the impact speed increases, the resulted
shock becomes stronger, then the influence of the accumulation of
the particles becomes stronger. If the strength of the material
close to the lower side of the cavity cannot withstand the
accumulation of particles, the upward speed of particles will become
larger and larger, and the cavity will collapse asymmetrically. The
upward speed of these particles in Fig.\ref{Fig7}(a) is about
$550\mathrm{m/s}$, increases to $844\mathrm{m/s}$ in
Fig.\ref{Fig7}(b), reaches $1000\mathrm{m/s}$ and a jet can be
observed in Fig.\ref{Fig7}(c), after that, these particles rush to
the upper side of cavity at a speed about $1000\mathrm{m/s}$. At the
time of Fig.\ref{Fig7}(d) the jet just impacts the upper side of
cavity, which causes of a distinct increase of pressure. The course
of collapsing of the lower cavity is similar.

\begin{figure}[btp]
\centering
\includegraphics*[width=0.47\textwidth]{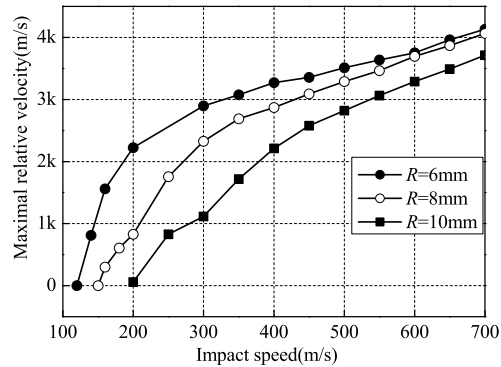}
\caption{The maximum relative velocity versus the initial impacting
speed. The radius of cavity is set as 6mm, 8mm and 10mm,
respectively.} \label{Fig_Velocity}
\end{figure}

\begin{figure}[btp]
\centering
\includegraphics*[width=0.47\textwidth]{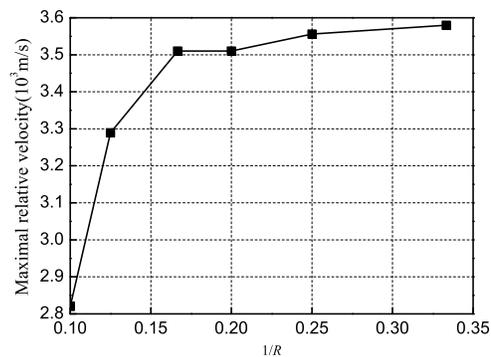}
\caption{The maximum relative velocity versus the radius of cavity.
The impact speed is set as 500m/s.} \label{Fig_ChangeR_V}
\end{figure}

To quantitatively describe the collapsing procedure of the shocked
cavity, we measure the maximum speed of particles being close to the
upper-steam wall relative to that being close to the down-stream
wall. Fig.\ref{Fig_Velocity} shows the maximum relative velocity
versus to the impacting speed. For the investigated cases, the
former increases nonlinearly with the impacting speed and approaches
to a constant; the smaller the initial radius, the larger the
maximum relative velocity. A specific case with an impacting speed
$v=500$ m/s is shown in Fig.\ref{Fig_ChangeR_V}. The approached
constant is about $3.58\mathrm{Km/s}$.

\begin{figure}[btp]
\centering
\includegraphics*[width=0.47\textwidth]{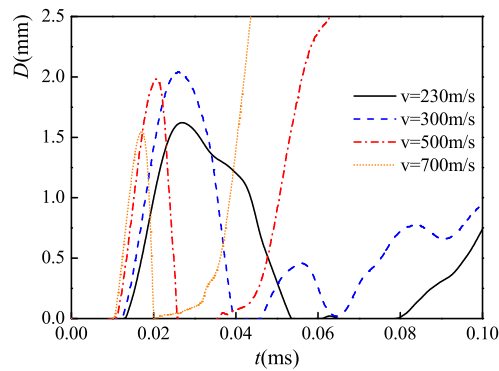}
\caption{(Color online) The separation distance versus time $t$ for
different impacting speeds with strong shock. Hear $D$ and has the
same meaning as Fig.\ref{Fig_R10_ChangeV}}
\label{Fig_V150_ChangeR_HighV}
\end{figure}

Also, the phenomena of the secondary impacting are observed in cases
the with strong shock. Fig.\ref{Fig_V150_ChangeR_HighV} shows the
effects of initial impacting speed $v$ on the separation distance
$D$ between the two impacting bodies. The initial impacting speed
$v$ is set as $230\mathrm{m/s}$, $300\mathrm{m/s}$,
$500\mathrm{m/s}$ and $700\mathrm{m/s}$, respectively.  Contrast to
the cases with weak shock, $D_{\max}$ doesn't increase monotonously
with the initial impact speed. When the initial impact speed is not
very large (lower than 300m/s), $D_{\max}$ increases
correspondingly; but when the impact speed is large (higher than
300m/s), $D_{\max}$ decreases. The critical case with $v=300$ m/s is
referred to Fig.\ref{Fig7}. A physical explanation based on the case
with $R=10$ mm is as below. When the impacting speed $v$ is lower
than 300 m/s (see Fig. \ref{Fig3}), the cavities shrink but without
jetting. The strength of the rarefactive wave increases with the
initial impacting speed $v$. When $v$ is larger than 300 m/s, the
cavities collapse and cave-in which dissipates a considerable amount
of the kinetic energy. The amount of the dissipated energy exceeds
the increasing of the total kinetic energy via increasing $v$.
Therefore, the strength of the rarefactive waves do not increase
with $v$ but decrease.

\subsection{On ``hot-spots''}
The reactivity of porous energetic material  depends greatly on the
nature of the ``hot-spots'' formed by shocks as they move through
the material. As observed in the gas-gun
experiments\cite{Sheffield1996}, coarse HMX produce ``hot-spots''
that are large enough to persist for a long time whereas fine
material seems to produce ``hot-spots'' that are smaller in size and
dissipate quickly. The threshold-to-initiation is the limit where
exothermic chemical energy release is balanced by energy dissipated
away from the ``hot-spot'' reaction. In the work of Baer, et
al\cite{Baer1998}, the mesoscale processes of consolidation,
deformation and reaction of shocked porous energetic materials are
studied using shock physics analysis of impact on a collection of
discrete HMX ``crystals''. In our work, we focus on the
``hot-spots'' produced in the shocked HMX material with cavities.

\begin{figure}[btp]
\centering
\caption{(See attached 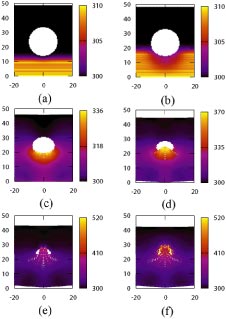) Snapshots for the impact of two
blocks with a single cavity in each. Impact velocity
$v=300\mathrm{m/s}$. From blue to red, the color in the figure shows
the increase of local temperature. The unit of temperature is
$\mathrm{K}$. (a) $t=0.004\mathrm{ms}$, (b) $t=0.006\mathrm{ms}$,
(c) $t=0.014\mathrm{ms}$, (d) $t=0.020\mathrm{ms}$, (e)
$t=0.025\mathrm{ms}$, (f) $t=0.030\mathrm{ms}$.} \label{Fig4}
\end{figure}

Fig.\ref{Fig4} shows the global procedure of the production of the
``hot-spots''. The impact speed is $300\mathrm{m/s}$, the initial
sizes of blocks are the same as those in section
\ref{sec-validation}, and the initial radius of every cavity is
$10\mathrm{mm}$. The initial temperature of the simulated system is
 $300\mathrm{K}$. In order to exhibit clearly, only the upper block
is shown. The procedure can be described as follows,

(a), The two blocks collide and the shocks begin to propagate
towards the cavities. The temperature in the shocked pure HMX is
about $310\mathrm{K}$(See in Fig.\ref{Fig4}(a)).

(b), The shocks reach the cavities, then rarefactive waves are
reflected back from the upstream boundaries of the cavities, the
temperatures in the regions which are first shocked and then
rarefacted by waves decrease a little (See in Fig.\ref{Fig4}(b)).

(c), The shocks continue to propagate. The sizes of the cavities
reduce as time goes on, the vertical deformation is more serious
than the horizontal one. Opposite to the last item, the temperatures
in the regions neighboring to the upstream sides of cavities
increase. The reason is that the plastic work increases continuously
(See in Fig.\ref{Fig4}(c)).

(d), A jet appears in each cavity, and the temperature of jet material increases
as time goes on(See in Fig.\ref{Fig4}(d)).

(e), The jet material impacts  the downstream side of the cavity,  which
cause a distinct increase of temperature to about $520\mathrm{K}$
(See in Fig.\ref{Fig4}(e)).

(f), The cavities continues to collapse until they are crammed, a
``hot-spot'' with the temperature about $520\mathrm{K}$ is produced
in each colliding region (See in Fig.\ref{Fig4}(f)).

\begin{figure}[btp]
\centering
\includegraphics*[width=0.47\textwidth]{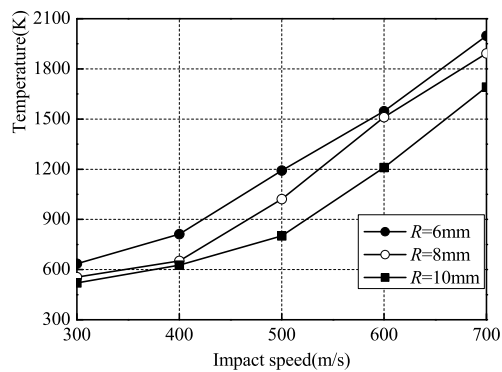}
\caption{The impact speed dependence of the temperature of the ``hot
spot''. The radius of cavities is set as 6mm, 8mm and 10mm,
respectively. The unit of temperature is $\mathrm{K}$.}
\label{Fig_Temperature}
\end{figure}

Fig.\ref{Fig_Temperature} shows the temperatures of ``hot-spot''
produced by different impact speeds for different sizes of cavities.
It can be seen that the temperature increases nearly parabolically
with the impacting speed $v$ for a fixed size of cavity, which is
consistent with the Hugoniot relation, $P=\rho_0 (c_0 + \lambda u)$,
where $u=v/2$ is the particle speed after shock. At the same time,
plastic work increases also nearly parabolically with the impacting
speed $v$.
 Also, it can be found that the
temperature of ``hot-spot'' produced in smaller cavity is higher
than that in larger ones with the same impact speed. As analyzed in
section \ref{sec-strongshock}, the maximum relative speed produced
by smaller cavities is larger than that by larger ones, which cause
higher temperature, for the investigated cases.

\section{Conclusion\label{SEC-CON}}

The dynamical response of high melting explosive with cavities under
shock is investigated by the generalized interpolation material
point method where the criterion for contacting is improved. An
elastic-to-plastic and thermal-dynamical model with the
Mie-Gr\"uneissen equation of state consulting the Rankine-Hugoniot
curve is used to simulate the mechanical and thermal behaviors. A
phenomena of secondary impacting is observed for the case of
colliding of two bodies with a single cavity in each. For weak
shocks, the cavities shrink in a nearly symmetric way, the phenomena
of secondary impacting becomes more distinctly as the impact speed
increases. But for strong shocks, the situation becomes complex
because the asymmetric collapse of cavity influences the reflection
of shock. The course for the collapse of cavity is studied for
various cavity sizes. For the checked cases, smaller cavities
collapse more easily. Our numerical results show that the existence
of cavities greatly help the creation of ``hot-spots'' with a much
higher temperature, which is very important for
 igniting of the explosive materials. The influence of
various factors, including impact speed, size of cavity, to the
temperature of ``hot-spots'' is investigated. The temperature of
``hot-spots'' increases nearly parabolically with the impact speed
for a fixed cavity size. The smaller the cavity, the higher the
temperature of ``hot-spots'', for cases investigated. If the cavity
is too small, the effects of cavity collapse becomes very weak.
Future work includes the incorporation of thermal diffusion, strain
rate effects, etc into the simulations, and cases with cavities
filled by gases.

\ack{ We warmly thank Shigang Chen, Haifeng Liu, Song Jiang,
Xingping Liu, Xijun Yu, Zhijun Shen, Yangjun Ying, Guoxi Ni, Yun Xu
 and Yingjun Li for helpful discussions. We acknowledge support by Science
Foundation of Laboratory of Computational Physics,  National Science
Foundation (Grant Nos.10702010 and 10775018) and National Basic
Research Program (Grant No. 2002BC412701) of China.}

\end{document}